\documentclass[10pt]{article}
\pdfoutput =1
\usepackage{graphicx}
\usepackage{amsmath,amssymb,amsthm,latexsym}
\usepackage{amsfonts} 
\usepackage{caption}
\usepackage{subfigure}
\usepackage{slashed}
\usepackage{float} 
\DeclareGraphicsExtensions{.pdf}

\begin{document}
\date{}
\title{{\bf{\Large AdS-CFT Correspondence in Noncommutative background, related thermodynamics and Holographic Superconductor in Magnetic Field}}}
\author{{\normalsize Souvik PRAMANIK}$
$\thanks{E-mail:  souvick.in@gmail.com}, 
{\normalsize Subir GHOSH}$
$\thanks{E-mail:  subir$_-$ghosh2@rediffmail.com}\\[0.5cm]
{\small{\emph{Physics and Applied Mathematics Unit,}}}\\
{\small{\emph{Indian Statistical Institute, }}}\\
{\small{\emph{203 B. T. Road, Kolkata 700108, India.}}}}


\maketitle
\begin{abstract}
In this work, we formulate a Non-Commutative (NC) extension of AdS-CFT correspondence that is manifested in the modification of behavior of a holographic superconductor. The noncommutativity is introduced in the model through the NC corrected AdS charged black hole metric developed by Nicolini,  Smailagic and Spallucci. First of all we discuss thermodynamic properties of this black hole in  Euclidean formalism.  In particular, we compute trace of the boundary energy-momentum tensor which, as expected, is non-zero due to the NC scale introduced in the model. Our findings indicate  that the non-commutative effects tend to work against the black hole hair formation. This, in turn, has an adverse effect on the holographic superconductor by making the superconducting phase more fragile. This is reflected in the reduced value of the critical magnetic field and critical temperature. Finally, we comment on  a qualitative agreement between our (holographic superconductor)  result and that obtained for a conventional superconductor in NC space in a purely condensed matter scenario. In both cases noncommutativity tends to oppose the superconducting phase.\\

\end{abstract}
\section{Introduction}
The AdS-CFT correspondence \cite{Phys.Rept.323.183.2000} has the capability to address the issues in strongly interacting systems in the context of Condensed Matter physics by exploiting results obtained in weakly coupled systems in the High Energy physics. A notable example of a strongly coupled Condensed Matter system  is the high $T_c$ superconductor, some of whose generic features have been recovered from weakly coupled gravity systems by exploiting the AdS-CFT correspondence. Indeed, in AdS-CFT, there exists explicit mapping between relevant operators and parameters of a field theory in the bulk of AdS space-time to those of a Conformal Field Theory living in the (one dimension lower) boundary. Gubser has shown that an Abelian Higgs model in AdS space can lead to a spontaneous symmetry breaking if one induce a scalar hair near the black hole horizon \cite{Phys.Rev.D.78.065034.2008}. AdS-CFT correspondence can show an interesting analogy with thin-film superconductors. In particular, the black hole Hawking temperature and hair is identified with the superconductor temperature and the condensate, respectively. A very simple example that can generate necessary results of a holographic superconductor \cite{hart} is the so called probe limit where the Maxwell field and the scalar field do not generate back reactions on the metric. With this limit one can consider the effect of Schwarzschild-AdS metric on scalar and Maxwell fields, instead of taking the background metric to be Reissner-Nordstrom in AdS. Though  afterwards the Reissner-Nordstrom metric has also been studied in \cite{hart1} (for review see \cite{holo}) which is a more general scenario.

In order to study the effects of a new property (either in the AdS or in the CFT sector), there are two basic approaches depending on ones interest and information content; one is top-down and another is a bottom-up approach. If the string theory realization of the new model in the bulk is known the former is preferable since one has a handle on the boundary theory via the AdS-CFT dictionary. On the other hand, if the stringy origin of a phenomenologically interesting bulk theory is not clear, the latter option is advisable where one's primary interest is to see whether the new bulk effect reveals a novel feature  of the boundary theory. Extension of theories in Non-Commutative (NC) space (or spacetime) is a topical area of interest, essentially in high energy physics context and there is a vast literature dealing with generalizations of field theories in NC space (or spacetime). (For reviews, see \cite{nft}). Here, generically NC effects depend on a new NC parameter $\theta$ that is imposed from outside in an ad-hoc way. On the other hand, NC effects have recently become very interesting in condensed matter physics \cite{xiao}, (such as Berry curvature effects, Landau problem, to name a few),  where, on the contrary, NC generalizations are interpreted as effective theories and NC parameter is not a new parameter, but is constructed from the system itself (as for example Berry curvature, that will be relevant in our case, from external magnetic field). In the present work we follow the bottom-up approach since we consider a specific form of NC extension of Schwarzschild-AdS black hole, provided by Nicolini, Smailagic and Spalucci \cite{nicolini}, in the bulk theory whose string theoretic origin is not well understood {\footnote{In this context, let us mention that, another well-known form of NC spacetime, revealed in the work of Seiberg and Witten \cite{string} is a direct descendant of string theory and can play a significant role in a top-down approach. We have not considered this scenario in the present paper.}}. Quite interestingly, we have been able to show that there is a qualitative matching between basic features of NC-extended holographic superconductors, obtained here for the first time, and NC-generalization of conventional superconductors studied in \cite{twist} in purely condensed matter scenario. 

In the present paper we extend the work initiated in our previous Letter \cite{sou} in that we now include effects of an external magnetic field on the Non-Commutativity (NC) induced generalized AdS-CFT correspondence proposed in \cite{sou}. Specifically,  our aim is to study the effects of Non-Commutative (NC) geometry on AdS-CFT correspondence and subsequently on the properties of holographic superconductors, in the presence of both bulk electric and magnetic fields. In this context, the backreaction of the matter field on the metric is assumed to be suppressed by considering the probe limit.

In Condensed Matter scenario magnetic field plays an important role since the behavior of a superconductor in external magnetic field distinguishes it being type I or type II. In general magnetic field is expelled out of the superconductor: the well known Meissner effect. This is captured by the following law,
\begin{equation}
	B_c(T)\approx B_c(0)\left[1-\left(\frac{T}{T_c} \right)^{2}  \right].
\end{equation}
The critical magnetic field $B_c(T)$ decreases as temperature $T$ gets closer to the critical temperature $T_c$ and vanishes for $T=T_c$ indicating the superconducting state ceases to exist beyond $T_c$. For temperatures lower than $T_c$ the system can withstand the external magnetic field up to $B_c(T)$. Thus the critical magnetic field $B_c$ turns out to be an important parameter   of the theory. When the external magnetic field exceeds $B_c$, it starts to penetrate the superconductor, but for type I the process is abrupt (a first order phase transition) whereas for type II the process is gradual with the formation of vortices of quantized flux (a second order phase transition). The high $T_c$ superconductors fall in the type II category. This feature is reproduced in the AdS-CFT analysis as well.

With the motivation of removing short distance singularities in quantum field theory, Snyder \cite{sny} for the  first time introduced a non-commutative feature in space-time in a covariant way. However, the idea did not gain popularity as it could not account for the radiative effects that were successfully explained by renormalized quantum field theory.  Later on, in a different context, NC field theory resurfaced in the seminal work of  by Seiberg and Witten \cite{string}. It was  shown that in the presence of an external two-form anti-symmetric $B_{\mu\nu}$ field, the $D$-branes on which open strings terminate, become  noncommutative in certain specific  low energy limit. The effects of a noncommutative space-time are incorporated in a field theory  prescribed in   \cite{string},  where normal products between local fields are required to be replaced by $*$-products. This allows one to study   NC QFTs in a conventional  perturbative framework for small NC parameter $\theta$ \cite{jac}. (However, this scheme is also plagued by the so called Ultraviolet-Infrared mixing problem.) On the other hand, for a consistent extension of a gauge theory an additional rule, known as the Seiberg-Witten map \cite{string} needs to be invoked. (See for example   \cite{nft} for a review). The most important outcome of NC space-time in NC field theory is the introduction of an independent length (or energy) scale.

In this perspective,  Nicolini, Smailagic and Spalucci \cite{nicolini}, in a recent work, have given a new way of inserting a length scale in a field theory where the conventional concept of point source is replaced by a smeared matter source, with  a Gaussian distribution (as an example). The  Einstein equation is directly solved in this background, thereby generating an extended  Schwarzschild metric, where  the NC effect appears through the minimum length scale $\theta$, induced by the Gaussian structure. Sometimes $\theta $ is identified as the Planck length. The main feature of this model is that the black hole singularity was successfully removed in this scenario.   Hawking radiation has been studied by exploiting this $\theta$-extended Schwarzschild metric  \cite{banerjee}. in \cite{nico2}, Hawking-Page crossover in  the such NC black hole metric in AdS has been discussed.  Thermalization processes  in black hole backgrounds with noncommutative corrections   have been studied in \cite{zeng}. 

In the present paper, we will show that this NC model satisfies the thermodynamical properties of a black hole. It has already been observed  that this model has the capability to avoid the singularities of a black hole which  plays a crucial role in the final stage of black hole evaporation. Therefore, it will be quite interesting to introduce this NC model in the context of the AdS-CFT correspondence and study the changes in the boundary CFT, that can lead to  measurable quantities to look for signatures of NC spacetime. 

In the present paper, we will study the bulk NC effect on Holographic Superconductors in the probe limit approximation. Our computations closely follow the analytical framework of \cite{greg} as exploited by \cite{dibakar}. It should be noted  that, to the lowest non-trivial order of $\theta$),  asymptotic behaviors of bulk fields are not modified by NC effects in a  qualitative way. This means that,  although the numerical parameters receive  NC corrections, functional forms of bulk fields remain unchanged. Hence we exploit the existing AdS-CFT dictionary  to calculate  $\theta$-effects in  critical temperature and  charge density relation of the Holographic Superconductor and subsequently the condensate-temperature relation. Our results pertain to the probe limit domain and reduce to the conventional results if the NC effects are removed, although the last part is achieved by a tricky limiting procedure.

The paper is organized as follows: In Section 2 we provide the NC-AdS black hole metric from the action for an Abelian gauge field coupled with a Higgs scalar in this NC background.  The thermodynamic aspects of NC AdS charged black holes are discussed. In Section 3, properties of the boundary energy momentum tensor are discussed. In Section 4, we obtain the field equations and we study the asymptotic behavior of the gauge and scalar fields. In Section 5 we  analyze the relation between the critical temperature and the charge density, and subsequently obtain the expression of the condensation operator. Afterward, in Section 6 we introduce an external magnetic field in the bulk and find the expression of the critical magnetic field for the NC holographic superconductor. We also comment on similar behavior of superconductors, residing in NC space, studied in the context of pure condensed matter physics. Finally, we conclude  in Section 7 by summarizing our results and mentioning some open problems.

\section{Non-commutative  $AdS$ Black Holes and related thermodynamics}
In the present section we recapitulate earlier results for the black hole in the NC AdS background (see for example \cite{ncb}). This is a generalization of the AdS Reissner-Nordstrom black hole in NC space.  This also helps us to fix our notations recall   the results used in our previous paper \cite{sou}. We start with the conventional action integral in the AdS space-time background given by
\begin{equation}
	I = \int d^4 x \sqrt{-g} \left[ \frac{1}{16 \pi G} \left( R+\frac{6}{L^2}\right)-\frac{1}{4}{F}_{\mu\nu}
	{F}^{\mu\nu}\right] + I_M, \label{action},
\end{equation}
where $L$ is the radius of AdS, ${F}_{\mu\nu}$ the Maxwell field in NC background and $I_M$ is the action of matter source. Throughout in this paper, we work in the system of natural unit with $c=\hbar=k_B=1$.

In NC background  the black hole is considered to have matter and charge density in the form of  a Gaussian distribution as proposed in \cite{nicolini} and for the ansatz  $A_\mu = (A_t, 0 , 0 , 0)$, a variation of the above action (\ref{action}) with respect to the metric tensor $g^{\mu\nu}$ provides the charged black hole solution in AdS as \cite{nicolini}
\begin{eqnarray}
	ds^2 &=& -f_1(r)dt^2+\frac{dr^2}{f_1(r)}+r^2 d\Omega^2, \nonumber\\
	f_1(r) &=& \kappa-\frac{4 M G}{r \sqrt{\pi}} \gamma\left(\frac{3}{2}, \frac{r^2}{4 \theta}\right)+\frac{G Q^2 }{\pi r^2} \left[\gamma^2 \left(\frac{1}{2}, \frac{r^2}{4 \theta}\right) - \frac{r}{\sqrt{2 \theta}} \gamma\left(\frac{1}{2},\frac{r^2}{2 \theta}\right) + \sqrt{\frac{2}{\theta}} r \ \gamma\left(\frac{3}{2}, \frac{r^2}{4 \theta}\right)\right]+\frac{r^2}{L^2}. \label{full-metric}
\end{eqnarray}
where, $\gamma(s, x) = \int_0^x t^{s-1} e^{-t} dt$ is the lower incomplete Gamma function and $\kappa$ the curvature which can take the values $0,+1,-1$ according to planar, spherical and hyperbolic space-time, respectively. Since, in this paper, we are intended to study the planar AdS Reissner-Nordstrom black hole we take $\kappa=0$. It can be noted that as $\theta \rightarrow 0$, the above metric (\ref{full-metric}) reduces to the usual one in AdS background.

However, in order to study the AdS-CFT correspondence the action is augmented by a charged scalar (Higgs) terms and to study the thermodynamical properties properly we have to introduce some boundary terms. Hence our complete action is
\begin{eqnarray}
	I &=& \int d^4 x \sqrt{-g} \left[ \frac{1}{16 \pi G} \left( R+\frac{6}{L^2}\right)-\frac{1}{4}{F}_{\mu\nu}
	{F}^{\mu\nu}-|\partial_\mu \Psi-i q{A}_\mu \Psi|^2-m^2|\Psi|^2\right] + \frac{1}{8 \pi G} \oint_{\partial M} d^3 x \sqrt{-\gamma} K \nonumber\\
	&& - \frac{1}{8 \pi G} \oint_{\partial M} d^3 x ~ L_{ct} (\gamma_{\mu\nu}) + I_m , \label{full-action}
\end{eqnarray}
where, in the second integral $K$ is the trace of the extrinsic curvature tensor $K_{\mu\nu}$ with respect to the boundary metric $\gamma_{\mu\nu}$ and the third integral $L_{ct}$ is the Lagrangian to encounter the divergence at AdS boundary.

Before proceeding further some comments about this action are in order. We follow the approach of \cite{kraus,york}. The $K$-boundary term is added in order to pose a well-defined variation problem. The counter term operates at the boundary to kill the divergence that appears in the stress tensor as the surface approaches the AdS boundary \cite{york}. A variation of the above action with respect to $g^{\mu\nu}$ provides the same metric (\ref{full-metric}) by considering $\psi$ is too small to back-react significantly upon the geometry \cite{Phys.Rev.D.78.065034.2008}. Since, we are intended to study large black holes, in the sense that the radius of the horizon is large enough compared to the NC length scale, then  $r^2 >> 4\theta$ and we can further approximate the Gamma functions, so that, the metric becomes
\begin{eqnarray}
	ds^2 &=& -f_1(r)dt^2+\frac{dr^2}{f_1(r)}+\frac{r^2}{L^2} (dx^2+dy^2),\nonumber\\
	f_1(r) &=& - \frac{2 M G}{r} + \frac{r^2}{L^2} + \frac{2 M G}{\sqrt{\pi \theta}} \ e^{-r^2 / 4 \theta} + \frac{G Q^2}{r^2} \left[1-\left(\frac{4 \sqrt{\theta}}{\sqrt{\pi} r} +\frac{r^2}{\sqrt{2} \pi \theta}\right) e^{-\frac{r^2}{4\theta}} \right] \label{appro-metric}
\end{eqnarray} 
Once again one can see that if the NC parameter $\theta \rightarrow 0$, the usual case will reproduce. The horizon radius can be found from $f_1(r_+) =0$, which provides a biquadratic equation of the form 
\begin{eqnarray}
	r_+^4 = (2 G M L^2) r_+ - G Q^2 L^2 -\left[ \frac{2 G M L^2}{\sqrt{\pi\theta}} r_+^2 - G Q^2 L^2 \left(\frac{4 \sqrt{\theta}}{\sqrt{\pi} r_+} +\frac{r_+^2}{\sqrt{2} \pi \theta}\right) \right]  e^{-\frac{r_+^2}{4\theta}} \label{hor-eqn}
\end{eqnarray} 
In order to obtain the NC effect properly, without complexity in calculations, it is quite tricky to study the near extremal black hole. First, we find the extremal configuration of the non-$\theta$ part, which is $r_0 = (\frac{G M L^2}{2})^{1/3}$ with the condition $G Q^2 L^2 = 3 (\frac{G M L^2}{2})^{4/3}$. Therefore, we can consider the solution of (\ref{hor-eqn}) as $r_+ = r_0 + \alpha$, where $\alpha$ correspond to the small NC correction terms. Substituting back this into (\ref{hor-eqn}) and considering the terms up to 2nd order of $\alpha$ we have that
\footnote{Note that, in this case we have to take care of terms up to 2nd order of $\alpha$, since the non-$\theta$ 1st order terms of $\alpha$ canceled out each other.}
\begin{eqnarray}
	r_\pm &=& r_0 \pm \sqrt{k(r_0)} e^{-\frac{r_0^2}{8 \theta}} - h(r_0) e^{-\frac{r_0^2}{4 \theta}}  \label{hor-radius}
\end{eqnarray}
where $k(r_0)= \left( \frac{2 \sqrt{\theta} r_0}{\sqrt{\pi}} - \frac{2  r_0^3}{ 3 \sqrt{\pi \theta}} + \frac{r_0^4}{ 2\sqrt{2} \pi \theta} \right)$, $ h(r_0) = \left( \frac{\sqrt{\pi}}{ \sqrt{ \theta}} + \frac{7  r_0^2}{ 6 \sqrt{\pi \theta}} - \frac{r_0^3}{ 2\sqrt{2} \pi \theta} - \frac{r_0^4}{ 6 \sqrt{\pi} \theta^{3/2}} + \frac{r_0^5}{ 8\sqrt{2} \pi \theta^2} \right)$. It is crucial to note here that a new exponential order which is the square root of our small exponential order $e^{-\frac{r_0^2}{4 \theta}}$ arises in the expression of $r_+$. Those terms are indeed larger than our small exponential order, and so from now on we have to deal with two small exponential orders.   

In order to find the Hawking temperature $T_H$, we can consider the radial distance just outside the horizon as $r= r_+ + \rho^2$, where $\rho^2$ is a very small quantity. Substituting this into the above metric (\ref{appro-metric}) and neglecting the products of $\rho^2$ and the exponential orders, we obtain
\begin{equation}
	ds^2 = \frac{4 L^2 r_+^3}{2 r_+^4 + (2 G M L^2) r_+ - 2 G Q^2 L^2} \left[ \left( \frac{2 r_+^4 + (2 G M L^2) r_+ - 2 G Q^2 L^2}{L^2 r_+^3} \right)^2 \rho^2 d\tau^2 + d\rho^2 \right] + r_+^2 d\Omega^2
\end{equation}  
where the metric has been made to be Euclidean by introducing the proper time $\tau$, given by $\tau = i t$. In order to avoid the singularity of the above metric, the proper time $\tau$ has to be periodic with period $\beta_* = \frac{4\pi L^2 r_+^3}{2 r_+^4 + (2 G M L^2) r_+ - 2 G Q^2 L^2}$. This expression provides the Hawking temperature $T_H$ as
\begin{equation}
	T_H^{-1} (r_0) \equiv \beta_* (r_0) = \frac{4 L^2 r_+^3}{2 r_+^4 + (2 G M L^2) r_+ - 2 G Q^2 L^2} = \frac{\pi L^2}{3[ \sqrt{k(r_0)} e^{-\frac{r_0^2}{8 \theta}} - h(r_0) e^{-\frac{r_0^2}{4 \theta}} ]} \label{hok-tem}
\end{equation}
The temperature at any radial distance $r$ can be found using the Hawking temperature as:
\begin{equation}
	T^{-1} (r,r_0) = \beta (r,r_0) = \int_0^{\beta_*} \sqrt{g_{tt}} dt = {\beta_*}(r_0) \times \sqrt{f_1(r)} \label{}
\end{equation}

In order to obtain the extrinsic curvature in (\ref{full-action}) at any finite distance $r$, we can write the metric as: $ds^2 = \frac{dr^2}{f_1(r)} + \gamma_{\mu\nu} dx^\mu ~ dx^\nu $, where $\gamma_{\mu\nu}$ is the extrinsic three-metric, given by, $\gamma_{\tau\tau}= f_1(r)$, $\gamma_{\theta\theta} = \frac{r^2}{L^2}$, $\gamma_{\phi\phi} = \frac{r^2 \sin^2\theta}{L^2}$. On the other hand, the extrinsic curvature tensor is defined by
\begin{equation}
	K^{\mu\nu} = - \frac{1}{2} (\nabla^\mu n^\nu + \nabla^\nu n^\mu) \label{ext-cur}
\end{equation}
where, $n^\mu$ is the unit normal vector along the $r$= constant surface. If we consider $n^\mu$ is along the radial direction, then it can be found to be $n^\mu = (0,\sqrt{f_1 (r)},0,0)$. So, using the $\gamma$'s and (\ref{ext-cur}) we find the trace of the extrinsic curvature tensor as $K = -\frac{1}{2 \sqrt{f_1 (r)}} \frac{\partial f_1(r)}{\partial r} - \frac{2}{r} \sqrt{f_1 (r)}$. Hence, the extrinsic curvature integral can be calculated to be 
\begin{equation}
	\frac{1}{8 \pi G} \oint_{\partial M} d^3 x \sqrt{\gamma} K = - \frac{\beta_*}{2 G L^2}\left[ 2 r f_1(r) + \frac{r^2}{2} \frac{\partial f_1(r)}{\partial r}\right]. \nonumber
\end{equation}

In order to find the counter term at the boundary of AdS we can approximate $f_1(r) \approx \frac{r^2}{L^2}$. Therefore the metric becomes: $ds^2 = \frac{ L^2 dr^2}{r^2} + \gamma_{\mu\nu} dx^\mu ~ dx^\nu $, where $\gamma_{\mu\nu}$ are given by $\gamma_{\tau\tau}= \frac{r^2}{L^2}, \gamma_{\theta\theta} = \frac{r^2}{L^2}, \gamma_{\phi\phi} = \frac{r^2 \sin^2\theta}{L^2}$. Using (\ref{ext-cur}), in this case we find $K = - \frac{3}{L}$, which provides $L_{ct} = - \sqrt{\gamma} \times \frac{3}{L}$. The corresponding integral can be calculated as $$- \frac{1}{8 \pi G} \oint_{\partial M} d^3 x ~ L_{ct} (\gamma_{\mu\nu}) = \frac{3 r^3 \beta}{2 G L^4}.$$

Generically, variation of the action produces a bulk term, that is proportional to the equations of motion, as well as a boundary term. Since we will be dealing with solutions of the dynamical equations the bulk term will vanish on-shell and only the surface term will contribute \cite{kraus}. This leads to the total action $I$ \cite{york}, 
\begin{equation}
	I = - \frac{\beta_*}{2 G L^2}\left[ 2 r f_1(r) + \frac{r^2}{2} \frac{\partial f_1(r)}{\partial r}\right] +  \frac{3 r^3 \beta}{2 G L^4}. \label{action-1}
\end{equation}
\begin{figure*}[htb]
	{\centering{\includegraphics[width = 7.5cm, height = 4.5 cm]{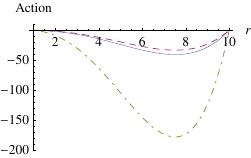}}
		\caption{Plot of total action $I$ against the radial coordinate $r$. In order to make the plot, we have considered $L=10$ and $r0=1$. Here, the solid, dashed and the dot-dashed curves are corresponding to $r_0^2/4\theta$ is equal to $1.5, 5, 10$. See that the action is negative for all values of $r \in [r0,L] $. This means that the black hole free energy is negative for all $r$, which implies thermodynamical consistent description of black-hole nucleation.}
		\label{fig-5}}
\end{figure*}
The important point about (\ref{action-1}) is that it is linear in $\beta$. This means that the subtraction term has no effect on the entropy, but on the normalization of the energy $E$. Henceforth, we can treat $r$ and $\beta$ as being independent variable. One can see from Figure-\ref{fig-5} that the action is negative for all values of $r$ between $r0$ to $L$ and for different values of NC parameter $\theta$. This means that the black hole free energy is negative, which implies thermodynamical consistent description of black-hole nucleation.  

In order to find the total energy $E$ within a fixed area $A= 4 \pi r^2$ we can use the expression of $I$. Note that, $I$ is a function of $r$ and $T$, so, the energy can be calculated as,
\begin{eqnarray}
	E \equiv \left[\frac{\partial I}{\partial \beta}\right]_A = - \frac{1}{2 G L^2} \frac{\partial \beta_*}{\partial r_0} \frac{\partial r_0}{\partial \beta}\left[ 2 r f_1 + \frac{r^2}{2} \frac{\partial f_1}{\partial r}\right]  - \frac{\beta_*}{2 G L^2}\left[ 2 r \frac{\partial f_1}{\partial r_0} \frac{\partial r_0}{\partial \beta} + \frac{r^2}{2} \frac{\partial^2 f_1}{\partial r_0 \partial r} \frac{\partial r_0}{\partial \beta} \right] + \frac{3 r^3}{2 G L^4}.
\end{eqnarray}
The entropy $S$ within the same region $A= 4 \pi r^2$ is computed as well,
\begin{eqnarray}
	S &\equiv & \beta \left[\frac{\partial I}{\partial \beta}\right]_A - I \nonumber\\
	&=&  - \frac{\beta_* \sqrt{f_1}}{2 G L^2} \frac{\partial \beta_*}{\partial r_0} \frac{\partial r_0}{\partial \beta}\left[ 2 r f_1 + \frac{r^2}{2} \frac{\partial f_1}{\partial r}\right]  - \frac{\beta_*^2 \sqrt{f_1}}{2 G L^2}\left[ 2 r \frac{\partial f_1}{\partial r_0} \frac{\partial r_0}{\partial \beta} + \frac{r^2}{2} \frac{\partial^2 f_1}{\partial r_0 \partial r} \frac{\partial r_0}{\partial \beta} \right] + \frac{\beta_*}{2 G L^2} \left[ 2 r f_1 + \frac{r^2}{2} \frac{\partial f_1}{\partial r}\right].
\end{eqnarray}
\begin{figure*}[htb]
	{\centering
		\begin{subfigure}[]
			{\centering{\includegraphics[width = 7.5cm, height = 5.0 cm]{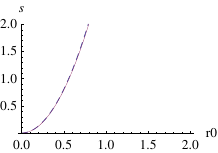}
					\label{fig-7-1}}}
		\end{subfigure}
		~
		\begin{subfigure}[]
			{\centering{\includegraphics[width = 7.5cm, height = 5.0 cm]{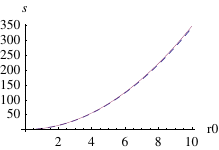}
					\label{fig-7-2}}}
		\end{subfigure}
		\caption{Plot of the entropy $4S$ against the black hole mass $M$ or $r_0$. Here, we have taken $L=10$ and $r=r_+$. In the first and the second figure we have considered $r_0^2/4\theta$ is equal to $1.5$, and $10$ respectively. Both curves are shown by the solid curves in the figure, whereas, the area of the event horizon $A = 4 \pi r_+^2$ is given by the dashed curves. From these plots, one can conclude that black-hole area law is consistent in this NC case.}
		\label{fig-7}}
\end{figure*}
From Figure-\ref{fig-7}, one can see that the entropy is an increasing function of black hole mass $M$ in this NC scenario. Since, the plot of the entropy coincides with the plot of horizon area $A$, so the black-hole area law is maintained in this NC scenario.

The thermodynamics of a black hole depend on the local stability in the canonical ensemble. This thermal stability is determined by the Heat capacity. The heat capacity $C_A$ is defined at a constant area $A = 4 \pi r^2$ of the cavity boundary by
\begin{eqnarray}
	C_A \equiv T \left[ \frac{\partial S}{\partial T}\right]_A &=& \left[ \frac{\partial E}{\partial T}\right]_A \nonumber\\
	&=& \frac{\beta_*^2 \sqrt{f_1}}{ 2 G L^2 \left( \frac{\partial \beta_*}{\partial r_0} + \frac{\beta_*}{2 f_1}\right)} \left[ \frac{\partial^2 \beta_*}{\partial r_0^2} \frac{\partial r_0}{\partial \beta}\left( 2 r f_1 + \frac{r^2}{2} \frac{\partial f_1}{\partial r}\right)  + 2 \frac{\partial \beta_*}{\partial r_0} \frac{\partial r_0}{\partial \beta} \left( 2 r \frac{\partial f_1}{\partial r_0} + \frac{r^2}{2} \frac{\partial^2 f_1}{\partial r_0 \partial r} \right) \right. \nonumber\\
	&& \left.+ \beta_* \frac{\partial r_0}{\partial \beta} \left( 2 r \frac{\partial ^2 f_1}{\partial r_0^2} + \frac{r^2}{2} \frac{\partial^3 f_1}{\partial^2 r_0 \partial r}\right) \right]
\end{eqnarray}
\begin{figure*}[htb]
	{\centering
		\begin{subfigure}[]
			{\centering{\includegraphics[width = 7.5cm, height = 5.0 cm]{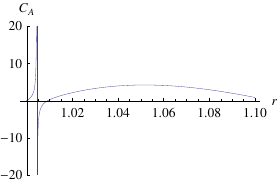}
					\label{fig-8-1}}}
		\end{subfigure}
		~
		\begin{subfigure}[]
			{\centering{\includegraphics[width = 7.5cm, height = 5.0 cm]{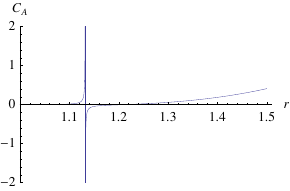}
					\label{fig-8-2}}}
	\end{subfigure}}
	\caption{Plot of the heat capacity $C_A$ against the radial coordinate $r$. Here, we have considered $L=10$ and $r0=1$. In the first and the second figure we have considered the NC parameter as $r_0^2/4\theta$ is equal to $1.5$, and $10$ respectively. We have plotted $r$ from the horizon radius $r_+$ to a finite distance less than the AdS radius $L$. In both the figure we can see that the heat capacity is positive just outside the horizon, which talks about the thermal stability of this NC system.}
	\label{fig-8}
\end{figure*}
From Figure-\ref{fig-8} one can see that the heat capacity is positive up to some values of $r$ just outside the event horizon. This feature of $C_A$ does not depend on which values of NC parameter have been taken into consideration. Hence, the NC black hole thermodynamics, here we are dealing with is stable. One can also check that the root-mean-square energy $(\Delta E)^2 = C_A T^2$ is a finite positive quantity as $r$ approaches to $r_+$. This means that no energy can be added to a black hole without changing its radius.

The "surface pressure" conjugate to the area $A = 4 \pi r^2$ is defined in analogy with the ordinary pressure by
\begin{eqnarray}
	\sigma &=& - \left[ \frac{\partial E}{\partial A}\right]_S \nonumber\\
	&=& \frac{1}{16 \pi r ~ G L^2} \left[ \frac{\partial \beta_*}{\partial r_0} \frac{\partial r_0}{\partial \beta} \left( 2 f_1 + 3 r \frac{\partial f_1}{\partial r} + \frac{r^2}{2} \frac{\partial^2 f_1}{\partial r_0 \partial r}\right) + \beta_* \left( 2 \frac{\partial f_1}{\partial r_0} \frac{\partial r_0}{\partial \beta} + 3 r \frac{\partial r_0}{\partial \beta} \frac{\partial^2 f_1}{\partial r_0 \partial r} + \frac{r^2}{2} \frac{\partial^3 f_1}{\partial r_0 \partial^2 r} \frac{\partial r_0}{\partial \beta} \right) \right] - \frac{9 r^2}{2 G L^4} \nonumber
\end{eqnarray} 
\begin{figure*}[htb]
	{\centering
		\begin{subfigure}[]
			{\centering{\includegraphics[width = 7.5cm, height = 4.5 cm]{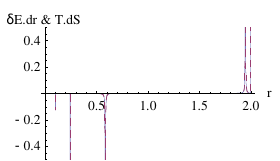}
					\label{pic-9-1}}}
		\end{subfigure}
		~
		\begin{subfigure}[]
			{\centering{\includegraphics[width = 7.5 cm, height = 4.5 cm]{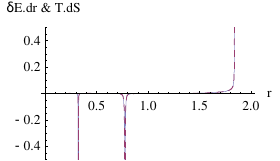}
					\label{pic-9-2}}}
	\end{subfigure}}
	\caption{Plot of $T dS$ and $\frac{\partial E}{\partial r_0} dr_0$ with $r_0$ for the different values of the non-commutative parameter given by $\theta= r_0^2/(4\times 1.5)$ and $r_0^2/(4\times 10)$, respectively. Here we have considered $L=2$ and $r=2$. The two plot shows $T dS$ and $\frac{\partial E}{\partial r_0} dr_0$ are same and independent of the choice of the NC parameter. }
	\label{fig-9}
\end{figure*}
\begin{figure*}[htb]
	{\centering
		\begin{subfigure}[]
			{\centering{\includegraphics[width = 7.5cm, height = 5.0 cm]{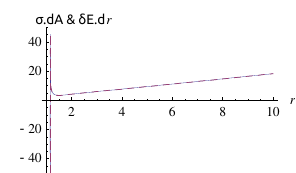}
					\label{pic-10-1}}}
		\end{subfigure}
		~
		\begin{subfigure}[]
			{\centering{\includegraphics[width = 7.5cm, height = 5.0 cm]{energy-dr-1}
					\label{pic-10-2}}}
	\end{subfigure}}
	\caption{Plot of $\sigma dA$ and $\frac{\partial E}{\partial r_0} dr_0$ for the two values of the NC parameter, given by $r_0^2/4\theta$ is equal to $1.5$ and $10$, respectively. Here we have considered $L=2$ and $r0=1$. The above two plots shows $\sigma dA$ and $\frac{\partial E}{\partial r_0} dr_0$ are same and independent of the choice of the NC parameter.}
	\label{fig-10}
\end{figure*}
We now check the thermodynamical identity given by $dE = T dS - \sigma dA$. Since, it is very difficult to show the satisfaction of this identity analytically in our system, we approach to the numerical simulations in this context. In the first and the second figure of Figure-\ref{fig-9}, $T dS$ and $\frac{\partial E}{\partial r_0} dr_0$ have been plotted with $r_0$ for the different values of the non-commutative parameter given by $\theta= r_0^2/(4\times 1.5)$ and $r_0^2/(4\times 10)$, respectively. Whereas, in the first and the second figure of Figure-\ref{fig-10}, $\sigma dA$ and $\frac{\partial E}{\partial r_0} dr_0$ have been plotted for the two values of the non-commutative parameter, given by $r_0^2/4\theta$ is equal to $1.5$ and $10$, respectively. From these plots, one can easily see that the above identity is fully satisfied in this NC case. Furthermore, one can  check that $\sigma \rightarrow 0$ on the AdS boundary. Therefore, one can recover the result $dr_0 = T_H dS$, with the approximation $f_1(r) = \frac{r^2}{L^2}$. 
 
\section{Boundary energy momentum  Tensor}
Let us now compute the boundary energy momentum tensor. Expectedly, it will not be conserved since an explicit mass scale has been introduced by way of the NC parameter $\theta$. This explicit breaking of conformal symmetry is present at all scales, but because of the smallness of the NC parameter its effect will be appreciable at high energy in the bulk. Our aim is to observe effects of this in the low energy boundary theory of holographic superconductor {\footnote{It needs to be pointed out that the AdS-CFT correspondence has been generalized to theories with more general backgrounds and in models without supersymmetry and conformal symmetry (see the review \cite{cai} and references therein).}}. Interestingly enough, a form of Ultraviolet-Infrared (UV-IR) relation in AdS-CFT correspondence, first highlighted by Susskind and Witten \cite{suss} and subsequently studied by Peet and Polchinski \cite{peet}, is responsible for the NC effect to percolate from the bulk to the boundary. In words, near the conformal boundary the IR behavior of the bulk theory dictates the UV physics of the boundary theory and vice versa. For example, the IR divergence of the bulk action near the boundary is dual to the UV divergence of the boundary field theory. Later, at the end of Section 6, we have presented a possible specific instance of this correspondence between boundary NC holographic superconductors and conventional superconductor in NC space.

In the asymptotic boundary (at fixed $r$) the metric (\ref{appro-metric}) can be written as
$ds^2 = \frac{dr^2}{f_1(r)} + \gamma_{\mu\nu}~ dx^\mu dx^\nu $, where, $\gamma_{\mu\nu}$ are the boundary metric given by $\gamma_{t t}= - f_1(r)$, $\gamma_{\theta\theta} = \frac{r^2}{L^2}$, $\gamma_{\phi\phi} = \frac{r^2 \sin^2\theta}{L^2}$. Using the extrinsic curvature tensor is defined in (\ref{ext-cur}) and considering the unit normal vector to the boundary of AdS is given by, $n^\mu = (0,\sqrt{f_1 (r)},0,0)$, we obtain the trace of the extrinsic curvature tensor as $K = -\frac{1}{2 \sqrt{f_1 (r)}} \frac{\partial f_1(r)}{\partial r} - \frac{2}{r} \sqrt{f_1 (r)}$. 

The quasilocal stress  tensor at the boundary is defined by
\begin{equation}
	T^{\mu\nu} = \frac{1}{8 \pi G}\left[K^{\mu\nu} - K \gamma^{\mu\nu} + \frac{2}{\sqrt{-\gamma}} \frac{\delta S_{ct}}{\delta \gamma_{\mu\nu}}\right]
\end{equation}
Since, in this 2+1 dimensional case the counter action is given by $S_{ct} = -\frac{2}{L} \int \sqrt{-\gamma}$, so, the components of the stress energy tensor are found to be
\begin{eqnarray}
	T^{tt} &=& \frac{1}{8 \pi G} \left[- \frac{2}{r \sqrt{f_1(r)}} - \frac{2}{L} \gamma^{tt}\right] \nonumber\\
	T^{\theta\theta} &=& \frac{1}{8 \pi G} \left[ \frac{L^2 \sqrt{f_1(r)}}{r^3}+ \frac{L^2}{2 r^2 \sqrt{f_1(r)}}\frac{\partial f_1(r)}{\partial r} - \frac{2}{L} \gamma^{\theta\theta}\right] \nonumber\\
	T^{\phi\phi} &=& \frac{1}{8 \pi G} \left[ \frac{L^2 \sqrt{f_1(r)}}{r^3 \sin^2[\theta]}+ \frac{L^2}{2 r^2 \sqrt{f_1(r)}\sin^2[\theta]}\frac{\partial f_1(r)}{\partial r} - \frac{2}{L} \gamma^{\theta\theta}\right]
\end{eqnarray}
Hence, the trace of the energy momentum tensor (or stress tensor) is given by, 
\begin{eqnarray}
	T &=& \frac{1}{8 \pi G} \left[-2 K - \frac{6}{L}\right]\nonumber\\
	&=& \frac{1}{8 \pi G} \left[ \frac{ 4\sqrt{f_1(r)}}{r} + \frac{1}{ \sqrt{f_1(r)}}\frac{\partial f_1(r)}{\partial r} - \frac{6}{L}\right]
\end{eqnarray}
It is important to note here that the energy momentum tensor is no longer traceless, as expected, but it tends to zero asymptotically, since asymptotically, we have that $f_1(r) \approx \frac{r^2}{L^2}$ (as $r$ approaches to $L$ all the terms proportional to $1/r$ or higher order $1/r$ tend to zero. Since, we have the NC correction terms in the order of $1/r \times 1/e^{r^2}$, they decay more rapidly than others).

\section{ Asymptotic forms of scalar and gauge fields}
Before proceeding further into the actual computations using AdS-CFT correspondence an important point regarding our model and  exploiting the probe limit that we have mentioned earlier needs to be clarified. Recall that in the original work of Gubser although a charged black hole was considered the Schwarzchild-AdS metric was used, instead of  Reissner-Nordstrom-AdS. Following the same spirit, we have used the NC extension of Schwarzchild-AdS metric, instead of the NC extension of  Reissner-Nordstrom-AdS. Furthermore, one might wonder that more than one form of mater sources is involved here, the matter, the Maxwell as well as a scalar field, so one can study the NC contribution from all of them are taken into account. For details we refer to the review by Nicolini in \cite{nicolini} but the essential point is explained below.

It needs to be stressed that there is a fundamental difference between the methodology of introducing NC effects in a field theory: in the Seiberg-Witten form of NC, {\it{all the local product terms between the fields in the action is replaced by $*$-products and NC effects enter via $*$-product}}. But, as has been convincingly argued in \cite{nicolini}, the NC deformed metric in the $*$-product framework is not at all free from the singularity problem, for which the NC was introduced in the first place. On the contrary, the smeared source paradigm of \cite{nicolini}, that we are following here, is  indeed capable of removing the metric singularity. The underlying philosophy is that {\it{only the source terms, which in the present case are the matter and the $U(1)$ charge, needs to be replaced by smeared source terms}}. From the results presented in \cite{ncads,nicolini} it is clear that the charge dependent terms will die out faster, (just as in the conventional Reissner-Nordstrom-AdS case). Hence we conclude that NC effects from all types of matter sources are taken into account in the present model.

In this paper, we will concentrate on large black holes, i.e. for which $r_+^2 / 4\theta \gg 1$, where $r_+$ is the radius of the horizon. Outside the horizon, we can therefore consider $r^2 / 4 \theta \gg 1$, which approximates the gamma function to give an exponential function. Since we are interested in studying the asymptotic behavior of AdS, and restrict ourselves to the probe limit,  the $Q^2$-dependent back reaction terms are not considered further in  $f_1(r)$. Therefore, in this case, the metric is restricted to the form,
\begin{eqnarray}
	ds^2 &=& -f_1(r)dt^2+\frac{dr^2}{f_1(r)}+\frac{r^2}{L^2} (dx^2+dy^2),\nonumber\\
	f_1(r) &=& K - \frac{2 M G}{r} + \frac{r^2}{L^2} + \frac{2 M G}{\sqrt{\pi \theta}} \ e^{-r^2 / 4 \theta} \label{metric-2}
\end{eqnarray}
Thus NC corrections will enter only through $f_1(r)$.
Now, variations of the above action (\ref{action}) with respect to the scalar field $\psi$ and the gauge field $A_t = \phi(r)$ provide the governing equation for our system as
\begin{eqnarray}
	&& \psi'' + \left(\frac{f_1'(r)}{f_1(r)} + \frac{2}{r}\right)\psi'-\frac{m^2}{f_1(r)}
	\psi+\frac{ \phi^2}{f_1^2(r)}\psi = 0. \label{diff-si} \\
	&& \phi'' + \frac{2}{r}\phi' - \frac{2|\psi|^2}{f_1(r)}\phi = 0. \label{diff-fi}
\end{eqnarray}

Asymptotically, as $r \rightarrow \infty $ the term $ \frac{2 G M}{\sqrt{\pi \theta}} e^{-r^2 / 4 \theta} \rightarrow 0$. Thus, the metric coefficient $f_1(r)$ in (\ref{metric-2}) can be approximated as $\frac{1}{f_1(r)}\approx\frac{L^2}{r^2}$ (considering terms up to $\frac{1}{r^2}$
only). With this approximation, the asymptotic behavior of the scalar field $\psi$ can be found to be
\begin{equation}
	\psi = \frac{\psi^-}{r} + \frac{\psi^+}{r^2}, \label{si-asym}
\end{equation}
Here, $\psi^-$ and $\psi^+$ are constants and can be considered alternatively as a source and the value of the condensate $\langle \mathcal{O}\rangle$, respectively. In the article, we consider $\psi^- = 0$ and $\psi^+ \simeq \langle \mathcal{O}_2 \rangle $, i.e. there will be condensation without any source term.

Using the same approximation $\frac{1}{f_1(r)}\approx\frac{L^2}{r^2}$, the asymptotic behavior of the gauge field $\phi$ turns out to be
\begin{equation}
	\phi = \mu-\frac{\rho}{r} \label{fi-asyem}
\end{equation}
where $\mu$ and $\rho$ are constants and can be considered respectively as the chemical potential and the charge density.

Notice that, the effects of non-commutativity are not present in both the above asymptotic behavior of the scalar fields \cite{sou}. However, later on we will show that the effects of non-commutativity will appear through the near horizon solutions of $\psi$ and $\phi$.

\section{Condensation in Non-commutative background}
In this section we recover the results obtained in our previous paper \cite{sou} but now with a more straightforward analytical method as applied in \cite{greg, dibakar}. Changing the variable from $ r $ to $ z =\frac{r_+}{r} $ the above set of equations (\ref{diff-si},\ref{diff-fi}) turns out to be,
\begin{eqnarray}
	&& 	\psi''(z)+\frac{f^{'}(z)}{f(z)}\psi'(z) - \frac{m^2 r_+^{2}}{z^{4}f(z)} \psi(z) + \frac{r_+^{2}\phi^{2}(z)}{z^{4}f^{2}(z)} \psi(z) = 0, \label{diff-si-1} \\
	&& \phi''(z) - \frac{2  r_+^{2} \  \psi^2 (z)}{z^4 f(z)} \phi(z) =0. \label{diff-fi-1}
\end{eqnarray}

\textbf{Boundary conditions:}
\begin{itemize}
	\item As, $f(z)|_{z=1} = 0$, so, in order to make the quantity $g^{\mu\nu} A_\nu A_\nu$ finite and to obtain a regular solution of the scalar field $\psi$ at $z=1$, we have the horizon boundary conditions for $\phi$ and $\psi$ respectively as
	\begin{eqnarray}
		\phi (1)=0, \ \ \  \psi^{'}(1) = \frac{m^2 r_+^2}{f'(1)} \  \psi(1). \label{bd-1}
	\end{eqnarray}
\end{itemize}

With the above expressions (\ref{diff-si-1},\ref{diff-fi-1}) in hand, we now aim to derive an analytic expression for the critical temperature and the condensation values in non-commutative background. Since, all of $\phi$, $\psi$ and $f(z)$ are regular at the horizon, we can Taylor expand them near $z=1$, respectively as
\begin{eqnarray}
	&& \phi (z) = \phi (1) - \phi^{'}(1)(1-z) + \frac{1}{2}\phi^{''}(1)(1-z)^{2} + .. ..\label{tel-1} \\
	&& \psi(z) = \psi(1) - \psi^{'}(1)(1-z) + \frac{1}{2} \psi^{''}(1)(1-z)^{2} + ....\label{tel-2} \\
	&& f(z) = f(1) - f^{'}(1)(1-z) + \frac{1}{2} f^{''}(1)(1-z)^{2} + ....\label{tel-3}
\end{eqnarray}
Without loss of generality, we can choose $ \phi'(1)<0 $ and $ \psi(1)>0 $ \cite{dibakar}.

Using these Taylor expansions, near $ z=1 $ from (\ref{diff-fi-1}) we obtain,
\begin{equation}
	\phi^{''}(1) = - \frac{2 L^2 \psi^2 (1)\phi'(1)}{3} \left( 1 - \frac{\omega^{1/3}}{3 \sqrt{\pi \theta}} e^{-\frac{\omega^{2/3}}{ 4 \theta }} + \frac{\omega}{6 \sqrt{\pi} \ \theta^{3/2}} e^{-\frac{\omega^{2/3}}{ 4 \theta }} \right) \label{neq11},
\end{equation}
where  $ \omega =2G M L^2 $. Finally, substituting (\ref{neq11}) into (\ref{tel-1}) and using the first boundary condition of (\ref{bd-1}) we obtain,
\begin{equation}
	\phi(z)= - \phi'(1)(1-z) - \frac{L^2 \psi^2 (1)}{3} \left( 1 - \frac{\omega^{1/3}}{3 \sqrt{\pi \theta}} e^{-\frac{\omega^{2/3}}{ 4 \theta }} + \frac{\omega}{6 \sqrt{\pi} \theta^{3/2}} e^{-\frac{\omega^{2/3}}{ 4 \theta }} \right) \phi'(1)(1-z)^{2}  \label{fi-hor-sol}.
\end{equation}
Similarly, using the Taylor expansions, near $ z=1 $ from (\ref{diff-si-1}) we obtain,
\begin{eqnarray}
	\psi''(1) &=& \frac{8}{9}\psi(1) \left( 1 - \frac{2 \omega^{1/3}}{3 \sqrt{\pi \theta}} e^{-\frac{\omega^{2/3}}{ 4 \theta }} + \frac{7 \omega}{48 \sqrt{\pi} \ \theta^{3/2}} e^{-\frac{\omega^{2/3}}{ 4 \theta }} + \frac{\omega^{5/3}}{32 \sqrt{\pi} \ \theta^{5/2}} e^{-\frac{\omega^{2/3}}{ 4 \theta }} \right) \nonumber\\
	&& - \frac{ L^4 \psi(1) \phi'^{2}(1)}{18r_+^{2}} \left( 1 - \frac{2 \omega^{1/3}}{3 \sqrt{\pi \theta}} e^{-\frac{\omega^{2/3}}{ 4 \theta }} + \frac{\omega}{3 \sqrt{\pi} \ \theta^{3/2}} e^{-\frac{\omega^{2/3}}{ 4 \theta }} \right) \label{neq13}.
\end{eqnarray}
where we have used the second boundary condition of (\ref{bd-1}). Substituting (\ref{neq13}) into (\ref{tel-2}) and using the same boundary condition, we finally arrive at
\begin{eqnarray}
	\psi (z) &=& \frac{1}{3} \left( 1 + \frac{2\omega^{1/3}}{3 \sqrt{\pi \theta}} e^{-\frac{\omega^{2/3}}{ 4 \theta }} - \frac{\omega}{3 \sqrt{\pi} \ \theta^{3/2}} e^{-\frac{\omega^{2/3}}{ 4 \theta }} \right) \psi(1) +\frac{2}{3} \left( 1 - \frac{\omega^{1/3}}{3 \sqrt{\pi \theta}} e^{-\frac{\omega^{2/3}}{ 4 \theta }} + \frac{\omega}{6 \sqrt{\pi} \ \theta^{3/2}} e^{-\frac{\omega^{2/3}}{ 4 \theta }} \right) \psi(1) z \nonumber\\
	&& + \left[\frac{4}{9} \left( 1 - \frac{2 \omega^{1/3}}{3 \sqrt{\pi \theta}} e^{-\frac{\omega^{2/3}}{ 4 \theta }} + \frac{7 \omega}{48 \sqrt{\pi} \ \theta^{3/2}} e^{-\frac{\omega^{2/3}}{ 4 \theta }} + \frac{\omega^{5/3}}{32 \sqrt{\pi} \ \theta^{5/2}} e^{-\frac{\omega^{2/3}}{ 4 \theta }} \right) \right. \nonumber\\
	&& \left. - \frac{ L^4 \phi'^{2}(1)}{36 r_+^{2}} \left( 1 - \frac{2 \omega^{1/3}}{3 \sqrt{\pi \theta}} e^{-\frac{\omega^{2/3}}{ 4 \theta }} + \frac{\omega}{3 \sqrt{\pi} \ \theta^{3/2}} e^{-\frac{\omega^{2/3}}{ 4 \theta }} \right) \right] \psi(1) (1 - z)^2 \label{si-hor-sol}.
\end{eqnarray}

Following the methodology developed in \cite{greg,dibakar}, we can now match the solutions (\ref{fi-hor-sol} , \ref{si-hor-sol}) with the asymptotic behaviors (\ref{fi-asyem}, \ref{si-asym}) at some intermediate point $ z=z_0 $.

In order to match these two sets of asymptotic solutions smoothly at $ z=z_0 $, we have the following four conditions,
\begin{eqnarray}
	\mu -\frac{\rho z_0}{r_+} &=& - \phi'(1)(1-z_0) \left[1 + \frac{L^2 \psi^2 (1)}{3} \left( 1 - \frac{\omega^{1/3}}{3 \sqrt{\pi \theta}} e^{-\frac{\omega^{2/3}}{ 4 \theta }} + \frac{\omega}{6 \sqrt{\pi} \theta^{3/2}} e^{-\frac{\omega^{2/3}}{ 4 \theta }} \right) (1-z_0) \right], \nonumber \\
	-\frac{\rho }{r_+} &=&  \phi'(1) + \frac{2 L^2 \psi^2 (1)}{3} \left( 1 - \frac{\omega^{1/3}}{3 \sqrt{\pi \theta}} e^{-\frac{\omega^{2/3}}{ 4 \theta }} + \frac{\omega}{6 \sqrt{\pi} \theta^{3/2}} e^{-\frac{\omega^{2/3}}{ 4 \theta }} \right) \phi'(1)(1-z_0) \label{com-1},
\end{eqnarray}
and
\begin{eqnarray}
	J^+ z_0^{2} &=& \frac{1}{3} \left( 1 + \frac{2\omega^{1/3}}{3 \sqrt{\pi \theta}} e^{-\frac{\omega^{2/3}}{ 4 \theta }} - \frac{\omega}{3 \sqrt{\pi} \ \theta^{3/2}} e^{-\frac{\omega^{2/3}}{ 4 \theta }} \right) \psi(1) +\frac{2}{3} \left( 1 - \frac{\omega^{1/3}}{3 \sqrt{\pi \theta}} e^{-\frac{\omega^{2/3}}{ 4 \theta }} + \frac{\omega}{6 \sqrt{\pi} \ \theta^{3/2}} e^{-\frac{\omega^{2/3}}{ 4 \theta }} \right) \psi(1) z_0 \nonumber\\
	&& + \left[\frac{4}{9} \left( 1 - \frac{2 \omega^{1/3}}{3 \sqrt{\pi \theta}} e^{-\frac{\omega^{2/3}}{ 4 \theta }} + \frac{7 \omega}{48 \sqrt{\pi} \ \theta^{3/2}} e^{-\frac{\omega^{2/3}}{ 4 \theta }} + \frac{\omega^{5/3}}{32 \sqrt{\pi} \ \theta^{5/2}} e^{-\frac{\omega^{2/3}}{ 4 \theta }} \right) \right. \nonumber\\
	&& \left. - \frac{ L^4 \phi'^{2}(1)}{36 r_+^{2}} \left( 1 - \frac{2 \omega^{1/3}}{3 \sqrt{\pi \theta}} e^{-\frac{\omega^{2/3}}{ 4 \theta }} + \frac{\omega}{3 \sqrt{\pi} \ \theta^{3/2}} e^{-\frac{\omega^{2/3}}{ 4 \theta }} \right) \right] \psi(1) (1 - z_0)^2, \nonumber\\
	J^+ z_0 &=& \frac{1}{3} \left( 1 - \frac{\omega^{1/3}}{3 \sqrt{\pi \theta}} e^{-\frac{\omega^{2/3}}{ 4 \theta }} + \frac{\omega}{6 \sqrt{\pi} \ \theta^{3/2}} e^{-\frac{\omega^{2/3}}{ 4 \theta }} \right) \psi(1) - \left[\frac{4}{9} \left( 1 - \frac{2 \omega^{1/3}}{3 \sqrt{\pi \theta}} e^{-\frac{\omega^{2/3}}{ 4 \theta }} + \frac{7 \omega}{48 \sqrt{\pi} \ \theta^{3/2}} e^{-\frac{\omega^{2/3}}{ 4 \theta }} \right.\right. \nonumber\\
	&& \left.\left. + \frac{\omega^{5/3}}{32 \sqrt{\pi} \ \theta^{5/2}} e^{-\frac{\omega^{2/3}}{ 4 \theta }} \right) - \frac{ L^4 \phi'^{2}(1)}{36 r_+^{2}} \left( 1 - \frac{2 \omega^{1/3}}{3 \sqrt{\pi \theta}} e^{-\frac{\omega^{2/3}}{ 4 \theta }} + \frac{\omega}{3 \sqrt{\pi} \ \theta^{3/2}} e^{-\frac{\omega^{2/3}}{ 4 \theta }} \right) \right] \psi(1) (1 - z_0). \label{com-2}
\end{eqnarray}

\begin{figure*}[htb]
	{\centering
		\begin{subfigure}[]
			{\centering{\includegraphics[width = 6.5cm, height = 5.5 cm]{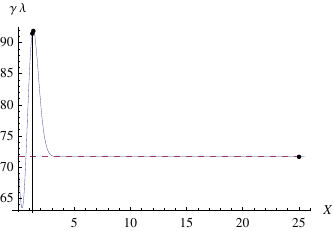}
					\label{pic-1-a}}}
		\end{subfigure}
		~
		\begin{subfigure}[]
			{\centering{\includegraphics[width = 6.0cm, height = 5.0 cm]{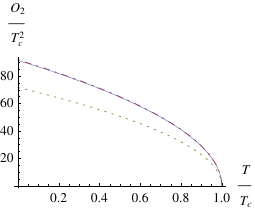}
					\label{pic-1-b}}}
	\end{subfigure}}
	\caption{In  figure \ref{pic-1-a} shows the plot of $\lambda $ vs. $\Lambda$.  In Figure \ref{pic-1-b}, condensation operator $O_2/T_c^2$ is plotted  against  black hole temperature $T$. Here, the continuous,  dotdashed and  dashed parabolas  correspond to the allowed boundary $\Lambda= 1.277$, the maximum magnitude $\Lambda= 1.366$ and for the normal case without non-commutativity, as well as the large values of $\Lambda$ ($>25.02$), respectively.}
	\label{fig-1}
\end{figure*}

From the second equation of (\ref{com-1}) we obtain,
\begin{equation}
	\psi^{2}(1) = \frac{3}{2 L^2 (1-z_0)} \left(\frac{\rho}{ r_+ \times [-\phi'(1)]} -1\right) \times \left( 1 + \frac{\omega^{1/3}}{3 \sqrt{\pi \theta}} e^{-\frac{\omega^{2/3}}{ 4 \theta }} - \frac{\omega}{6 \sqrt{\pi} \  \theta^{3/2}} e^{-\frac{\omega^{2/3}}{ 4 \theta }} \right) \label{si-sqr}.
\end{equation}
Since, in this non-commutative scenario, the Hawking temperature \cite{hawking} $T$ is related to the horizon radius $r_+$ by the relation
\begin{equation}
	T = \frac{3 r_+}{ 4 \pi L^2} \left( 1 + \frac{\omega^{1/3}}{3 \sqrt{\pi \theta}} e^{-\frac{\omega^{2/3}}{ 4 \theta }} - \frac{\omega}{6 \sqrt{\pi} \ \theta^{3/2}} e^{-\frac{\omega^{2/3}}{ 4 \theta }} \right) \label{T-hor-rel}
\end{equation}
we can write  (\ref{si-sqr}) in the form,
\begin{equation}
	\psi^{2} (1) =\frac{3}{2 L^2 (1-z_0)} \left(\frac{T_c}{T}\right)^{2}  \left(1-\frac{T^{2}}{T_c^{2}} \right) \times \left( 1 + \frac{\omega^{1/3}}{3 \sqrt{\pi \theta}} e^{-\frac{\omega^{2/3}}{ 4 \theta }} - \frac{\omega}{6 \sqrt{\pi} \  \theta^{3/2}} e^{-\frac{\omega^{2/3}}{ 4 \theta }} \right) \label{si-sqr-2},
\end{equation}
where, we have considered
\begin{equation}
	T_c = \frac{3\sqrt{\rho \ r_+}}{4\pi L^2 \sqrt{-\phi'(1)}}\left( 1 + \frac{\omega^{1/3}}{3 \sqrt{\pi \theta}} e^{-\frac{\omega^{2/3}}{ 4 \theta }} - \frac{\omega}{6 \sqrt{\pi} \ \theta^{3/2}} e^{-\frac{\omega^{2/3}}{ 4 \theta }} \right) \label{T-c}.
\end{equation}
Finally, for $ T\sim T_c $ i.e, very close to the critical temperature, from (\ref{si-sqr-2}) we obtain,
\begin{equation}
	\psi(1) = \sqrt{\frac{3}{ L^2 (1-z_0)}} \sqrt{1-\frac{T}{T_c}} \times \left( 1 + \frac{\omega^{1/3}}{6 \sqrt{\pi \theta}} e^{-\frac{\omega^{2/3}}{ 4 \theta }} - \frac{\omega}{12 \sqrt{\pi} \  \theta^{3/2}} e^{-\frac{\omega^{2/3}}{ 4 \theta }} \right) \label{si-1}.
\end{equation}
Again from (\ref{com-2}) we get,
\begin{eqnarray}
	J^+ &=& \frac{\psi (1) (2+z_0)}{3z_0} \left( 1 + \frac{(1 - z_0)}{3 (2 + z_0)} \frac{\omega^{1/3}}{\sqrt{\pi \theta}} e^{-\frac{\omega^{2/3}}{ 4 \theta }} - \frac{(1 - z_0)}{6 (2 + z_0)} \frac{\omega}{\sqrt{\pi} \  \theta^{3/2}} e^{-\frac{\omega^{2/3}}{ 4 \theta }} \right), \nonumber\\
	\beta &=& \frac{2}{L^2} \sqrt{\frac{7-z_0}{1-z_0}} \sqrt{ 1 + \left( \frac{(4 + 3 z_0)}{(7 - z_0)} \frac{\omega^{1/3}}{\sqrt{\pi \theta}} - \frac{(11 - z_0)}{4 (7 - z_0)} \frac{\omega}{\sqrt{\pi} \  \theta^{3/2}}  +  \frac{(1 - z_0)}{8 (7- z_0)} \frac{\omega^{5/3}}{\sqrt{\pi} \  \theta^{5/2}}\right) e^{-\frac{\omega^{2/3}}{ 4 \theta }} } \label{J-bta-exp}
\end{eqnarray}
where $\beta = \frac{[-\phi'(1)]}{r_+}$.
Finally, using (\ref{T-hor-rel}), (\ref{si-1}) and first equation of (\ref{J-bta-exp}), near the critical temperature ($ T\sim T_c $) the condensation operator is calculated as \cite{dibakar},
\begin{eqnarray}
	\langle \mathcal{O}_2\rangle &=& \sqrt{2} \  J^+ r_+^{2}\nonumber\\
	&=& \frac{16\sqrt{2} \pi^{2}}{9}\left(\frac{2+z_0}{3z_0} \right) \sqrt{\frac{3 L^6}{2 (1-z_0)}} \left[ 1 -  \frac{(4+ 5 z_0)}{6 (2 + z_0)} \left(\frac{\omega^{1/3}}{\sqrt{\pi \theta}} - \frac{\omega}{\sqrt{\pi} \  \theta^{3/2}}\right) e^{-\frac{\omega^{2/3}}{ 4 \theta }} \right] \ T_c^{2} \sqrt{1-\frac{T}{T_c}}. \label{o-2}
\end{eqnarray}
As we have mentioned before, NC effects have not brought about any qualitative change in the form of the condensation operator but has modified the numerical parameters. Rewriting (\ref{o-2}) as $\langle \mathcal{O}_2\rangle =\lambda  T_c^{2} \sqrt{1-\frac{T}{T_c}} $ we have plotted $\lambda $ against $\Lambda= \frac{(2 G M L^2)^{1/3}}{2 \sqrt{\theta}}$, a measure of the NC effect, in Figure \ref{pic-1-a}, whereas in Figure \ref{pic-1-b} we show the behavior of the dimensionless measure of the condensation operator $O_2/T_c^2$ against the dimensionless temperature $T/T_c$.

In  Figure \ref{pic-1-a}, the saturation value $\gamma$ for $\lambda$ indicated by the dashed line, represents the conventional value of $\lambda $ without NC effects. This continuous line shows that the coefficient of $T_c^{2} \sqrt{1-\frac{T}{T_c}}$ is independent of the black hole mass. On the other hand, the right portion of the vertical line (corresponding to $\Lambda= 1.277$) represents our allowed region $r_+^2 > 4 \theta$. One can see that, the NC coefficient $\lambda$ is maximum for $\Lambda= 1.366$, which is very near to the boundary of the allowed region and for large values of $\Lambda$ (when $\Lambda > 25.02$), $\lambda$ essentially stabilizes to the normal case indicating that NC effects seizes to be significant.

In Figure \ref{pic-1-b}, we have plotted the dimensionless condensation operator $O_2$ against the dimensionless measure of the black hole temperature $T$. Here, the continuum, the dotdashed and the dashed parabolas are corresponding to the allowed boundary $\Lambda= 1.277$, the maximum magnitude $\Lambda= 1.366$ and for the normal case, as well as the large values of $\Lambda$ ($>25.02$), respectively. From the figure we observe that in general NC effects lower the condensate value indicating that NC effects tend to destabilize the superconducting state.

\section{Condensation in presence of external magnetic field in non-commutative background}
	
The present section constitutes our main analysis where the effect of an external static magnetic field on a holographic superconductor has been studied by adding a magnetic field in the bulk. According to the gauge-gravity dictionary, this magnetic field asymptotically corresponds to a magnetic field $B({\bf{x}}) = F_{xy}({\bf{x}}, z\rightarrow 0)$ added to the boundary field theory. Again, since the condensate is small near the upper critical value ($ B_{c} $) of the magnetic field, therefore the scalar field $ \psi$ can be considered as a perturbation near the critical field strength $ B\sim B_{c} $. Hence, we adopt the following ansatz,
\begin{equation}
	A_t = \phi(z),~~~A_y = B x,~~~\Psi = \Psi (x,z)\label{neq24}.
\end{equation}
It is important to note here that the variation of the action (\ref{action}) with respect to metric tensor $g^{\mu\nu}$ with the above ansatz will give rise to additional terms in the expression of $f_1(z)$. However, since we are working on the probe limit,  these additional terms can be dropped and the expression of $f_1(z)$ remains same as (\ref{metric-2}).

With the above choice, the governing equation of $\phi$ remains same, whereas, the scalar field equation for $\Psi$ turns out to be,
\begin{eqnarray}
	\Psi''(x,z) + \frac{f_1^{'}(z)}{f_1(z)} \Psi'(x,z) - \frac{ m^2 r_+^{2}}{z^{4} f_1(z)} \Psi(x,z)  + \frac{r_+^{2}\phi^{2}(z)}{z^{4}f_1^{2}(z)} \Psi(x,z) + \frac{L^2}{z^{2}f_1(z)}[\partial_{x}^{2}\Psi(x,z) - B^{2}x^{2}\Psi (x,z)] = 0 \label{diff-si-3}.
\end{eqnarray}
In order to solve (\ref{diff-si-3}), we consider the separation of variable of the form $\Psi (x,z) = \psi(z) X(x)$. Substituting this form into (\ref{diff-si-3}) we finally get,
\begin{eqnarray}
	\frac{z^{2}f_1(z)}{L^2} \left[\frac{\psi''(z)}{\psi(z)}+ \frac{f_1'(z)}{f_1(z)}\frac{\psi'(z)}{\psi(z)}+\frac{\phi^{2}(z) r_+^{2}}{z^{4}f_1^{2}(z)}+\frac{2r_{+}^{2}}{z^{4}f_1(z)} \right]-\left[ -\frac{X''}{X} + B^{2} x^{2}\right]=0.
\end{eqnarray}

The equation for $ X(x)$ could be identified as the Schrodinger equation for a simple harmonic oscillator localized in one dimension with frequency determined by $ B $,
\begin{equation}
	-X^{''}(x) + B^{2}x^{2}X(x) = \lambda_{n} B X(x)
\end{equation}
where $ \lambda_{n}= 2n+1 $ denotes the separation constant. We consider the lowest mode ($ n=0 $) solution which is expected to be most stable \cite{hart,holo}.  As a quick comment that the NC effects considered here do not modify the observation presented in \cite{tamim} that increase in $B$ tends to shrink the size of the condensate.

With this particular choice, the equation for $ \psi(z) $  reduces to
\begin{equation}
	\psi''(z) + \frac{f_1^{'}(z)}{f_1(z)}\psi'(z) - \frac{ m^2 r_+^{2}}{z^{4}f_1(z)} \psi(z) + \frac{r_+^{2}\phi^{2}(z)}{z^{4}f_1^{2}(z)} \psi(z) = \frac{B L^2}{z^{2}f_1(z)} \psi(z) \label{diff-si-4}.
\end{equation}

\textbf{Boundary Condition:}

$\bullet$ In  NC scenario the regularity condition at the horizon $z=1 $  can be found from (\ref{diff-si-4}) as
\begin{equation}
	\psi'(1) = \left(\frac{m^2 r_+^2}{f_1'(1)} - \frac{B L^2}{f_1'(1)}\right) \psi(1) \label{bd-2}.
\end{equation}

$\bullet$ On the other hand, substituting the same approximate form $\frac{1}{f_1(r)}\approx\frac{L^2}{r^2}$ (considering terms up to $\frac{1}{r^2}$) into (\ref{diff-si-4}) we have the same asymptotic solution of the scalar field $\psi$ as,
\begin{equation}
	\psi(z) = J_- z + J_+ z^{2} \label{neq29}
\end{equation}
We again consider our previous choice $ J_- = 0 $.

Since, $\psi(z)$ is regular at the horizon, near $ z=1 $, we can again Taylor expand $ \psi(z) $ as (\ref{tel-2}). In order to obtain $ \psi'' (1) $ we substitute (\ref{tel-2}) and (\ref{tel-3}) into (\ref{diff-si-4}) which yields
\begin{eqnarray}
	\psi''(1) & =& \frac{8}{9} \left( 1 - \frac{2 \omega^{1/3}}{3 \sqrt{\pi \theta}} e^{-\frac{\omega^{2/3}}{ 4 \theta }} + \frac{7 \omega}{48 \sqrt{\pi} \ \theta^{3/2}} e^{-\frac{\omega^{2/3}}{ 4 \theta }} + \frac{\omega^{5/3}}{32 \sqrt{\pi} \ \theta^{5/2}} e^{-\frac{\omega^{2/3}}{ 4 \theta }} \right) \psi(1) \nonumber\\
	&&- \frac{ L^4 \phi'^{2}(1)}{18r_+^{2}} \left( 1 - \frac{2 \omega^{1/3}}{3 \sqrt{\pi \theta}} e^{-\frac{\omega^{2/3}}{ 4 \theta }} + \frac{\omega}{3 \sqrt{\pi} \ \theta^{3/2}} e^{-\frac{\omega^{2/3}}{ 4 \theta }} \right) \psi(1) \nonumber\\
	&& -\frac{5B L^4}{9 r_+^{2}} \left( 1 - \frac{2 \omega^{1/3}}{3 \sqrt{\pi \theta}} e^{-\frac{\omega^{2/3}}{ 4 \theta }} + \frac{11 \omega}{60 \sqrt{\pi} \ \theta^{3/2}} e^{-\frac{\omega^{2/3}}{ 4 \theta }} + \frac{\omega^{5/3}}{40 \sqrt{\pi} \ \theta^{5/2}} e^{-\frac{\omega^{2/3}}{ 4 \theta }} \right) \psi(1) \nonumber\\
	&& + \frac{B^{2} L^8}{18 r_+^{4}}  \left( 1 - \frac{2 \omega^{1/3}}{3 \sqrt{\pi \theta}} e^{-\frac{\omega^{2/3}}{ 4 \theta }} + \frac{\omega}{3 \sqrt{\pi} \ \theta^{3/2}} e^{-\frac{\omega^{2/3}}{ 4 \theta }} \right) \psi(1) \label{si''}.
\end{eqnarray}
Substituting (\ref{si''}) into (\ref{tel-2}) and using (\ref{bd-2}) we find,
\begin{eqnarray}
	\psi (z) & =& \frac{1}{3} \left( 1 + \frac{2\omega^{1/3}}{3 \sqrt{\pi \theta}} e^{-\frac{\omega^{2/3}}{ 4 \theta }} - \frac{\omega}{3 \sqrt{\pi} \ \theta^{3/2}} e^{-\frac{\omega^{2/3}}{ 4 \theta }} \right) \psi(1) + \frac{2}{3} \left( 1 - \frac{\omega^{1/3}}{3 \sqrt{\pi \theta}} e^{-\frac{\omega^{2/3}}{ 4 \theta }} + \frac{\omega}{6 \sqrt{\pi} \ \theta^{3/2}} e^{-\frac{\omega^{2/3}}{ 4 \theta }} \right) \psi(1) z \nonumber\\
	&& -\frac{B L^4}{3 r_+^{2}} \left( 1 - \frac{ \omega^{1/3}}{3 \sqrt{\pi \theta}} e^{-\frac{\omega^{2/3}}{ 4 \theta }} + \frac{\omega}{6 \sqrt{\pi} \ \theta^{3/2}} e^{-\frac{\omega^{2/3}}{ 4 \theta }} \right) \psi(1) (1 - z) \nonumber\\
	&& + \left[\frac{4}{9} \left( 1 - \frac{2 \omega^{1/3}}{3 \sqrt{\pi \theta}} e^{-\frac{\omega^{2/3}}{ 4 \theta }} + \frac{7 \omega}{48 \sqrt{\pi} \ \theta^{3/2}} e^{-\frac{\omega^{2/3}}{ 4 \theta }} + \frac{\omega^{5/3}}{32 \sqrt{\pi} \ \theta^{5/2}} e^{-\frac{\omega^{2/3}}{ 4 \theta }} \right) \right. \nonumber\\
	&& \left. - \frac{ L^4 \phi'^{2}(1)}{36 r_+^{2}} \left( 1 - \frac{2 \omega^{1/3}}{3 \sqrt{\pi \theta}} e^{-\frac{\omega^{2/3}}{ 4 \theta }} + \frac{\omega}{3 \sqrt{\pi} \ \theta^{3/2}} e^{-\frac{\omega^{2/3}}{ 4 \theta }} \right) \right. \nonumber\\
	&& \left. -\frac{5B L^4}{18 r_+^{2}} \left( 1 - \frac{2 \omega^{1/3}}{3 \sqrt{\pi \theta}} e^{-\frac{\omega^{2/3}}{ 4 \theta }} + \frac{11 \omega}{60 \sqrt{\pi} \ \theta^{3/2}} e^{-\frac{\omega^{2/3}}{ 4 \theta }} + \frac{\omega^{5/3}}{40 \sqrt{\pi} \ \theta^{5/2}} e^{-\frac{\omega^{2/3}}{ 4 \theta }} \right) \right. \nonumber\\
	&& \left. + \frac{B^{2} L^8}{36 r_+^{4}}  \left( 1 - \frac{2 \omega^{1/3}}{3 \sqrt{\pi \theta}} e^{-\frac{\omega^{2/3}}{ 4 \theta }} + \frac{\omega}{3 \sqrt{\pi} \ \theta^{3/2}} e^{-\frac{\omega^{2/3}}{ 4 \theta }} \right) \right] \psi(1) (1 - z)^2.
	\label{neq32}.
\end{eqnarray}

We can again match the above two solutions (\ref{neq29},\ref{neq32}) for some intermediate point $ z=z_0 $ and get two relations. If we eliminate $J_+$ from those relations, a quadratic equation for $B$ results:
\begin{eqnarray}
	&& B^{2} + 2 B \ \frac{r_+^2}{L^4} \frac{(1 + 2 z_0)}{(1 -z_0)} \left[ 1  + \left(\frac{(2 - z_0)}{(1+ 2 z_0)} \frac{\omega^{1/3}}{\sqrt{\pi \theta}} - \frac{(1 + z_0)}{4 (1 + 2 z_0)} \frac{\omega}{\sqrt{\pi} \  \theta^{3/2}} - \frac{(1 - z_0)}{8(1 + 2 z_0)} \frac{\omega^{5/3}}{\sqrt{\pi} \  \theta^{5/2}}\right) e^{-\frac{\omega^{2/3}}{ 4 \theta }} \right] \nonumber\\
	&& + \frac{4r_+^{4}}{L^8} \frac{(7 - z_0)}{(1 -z_0)} \left[ 1 + \left(\frac{(4 + z_0)}{(7- z_0)} \frac{\omega^{1/3}}{\sqrt{\pi \theta}} - \frac{(11- z_0)}{4 (7- z_0)} \frac{\omega}{\sqrt{\pi} \  \theta^{3/2}} + \frac{(1 - z_0)}{8(7-z_0)} \frac{\omega^{5/3}}{\sqrt{\pi} \  \theta^{5/2}}\right) e^{-\frac{\omega^{2/3}}{ 4 \theta }} \right] -  \frac{\phi'^{2}(1)r_+^{2}}{L^4} = 0, \nonumber\\
\end{eqnarray}
which has the solution,
\begin{eqnarray}
	B &=& \left\{ \frac{\phi'^{2}(1)r_+^{2}}{L^4} - \frac{9r_+^{4}(3-4z_0)}{ L^4 (1-z_0)^{2}} \left[1 + \left( \frac{2 (2 - 3 z_0)}{3 (3- 4 z_0)} \frac{\omega^{1/3}}{\sqrt{\pi \theta}} - \frac{(7- 9 z_0)}{6 (3 - 4 z_0)} \frac{\omega}{\sqrt{\pi} \  \theta^{3/2}} \right.\right.\right. \nonumber\\
	&& \left.\left.\left. + \frac{(1 - z_0)}{12 (3 - 4 z_0)} \frac{\omega^{5/3}}{\sqrt{\pi} \  \theta^{5/2}} \right) e^{-\frac{\omega^{2/3}}{ 4 \theta }} \right] \right\}^{1/2} \nonumber\\
	&& - \frac{r_+^{2} (1 + 2 z_0)}{L^4 (1-z_0)} \left[1 + \left( \frac{(2 - z_0)}{(1 + 2 z_0)} \frac{\omega^{1/3}}{\sqrt{\pi \theta}} - \frac{(1 + z_0)}{4 (1 + 2 z_0)} \frac{\omega}{\sqrt{\pi} \  \theta^{3/2}} - \frac{(1 - z_0)}{8 (1 + 2 z_0)} \frac{\omega^{5/3}}{\sqrt{\pi} \  \theta^{5/2}} \right) e^{-\frac{\omega^{2/3}}{ 4 \theta }} \right] \label{B}.
\end{eqnarray}

Now consider the case  where the value of the external magnetic field ($ B $) is very close to the upper critical value i.e, $ B\sim B_{c} $. This implies a vanishingly small condensation, so that, all the quadratic terms in $ \psi $ can be neglected. With this approximation from (\ref{diff-fi-1}) we obtain
\begin{equation}
	\phi''(z) = 0, \label{diff-fi-2}
\end{equation}
which has a solution, $\phi (z) = \frac{\rho}{r_+}(1-z)$. Near the asymptotic boundary ($ z\rightarrow 0 $) of the AdS, we can compare this solution with (\ref{fi-asyem}) and obtain
\begin{equation}
	\mu = \frac{\rho}{r_+} \label{rel-1}.
\end{equation}

On the other hand, near the horizon ($ z=1 $), from (\ref{diff-fi-2}) we have $\phi''(1) = 0$. Substituting this relation into the Taylor expansion of $\phi (z)$ near $z=1$, and using the boundary condition (\ref{bd-1}) we obtain
\begin{equation}
	\phi (z) = -\phi'(1)(1-z) \label{eq40}.
\end{equation}
Matching this solutions with (\ref{fi-asyem}) for any intermediate point $ z=z_0 $, we have
\begin{equation}
	\mu - \frac{\rho}{r_+} z_0 = -\phi'(1)(1-z_0) \label{rel-2}.
\end{equation}
From (\ref{rel-1}) and (\ref{rel-2}) we finally obtain
\begin{eqnarray}
	\phi'(1) = -\frac{\rho}{r_+} \label{neq-46}.
\end{eqnarray}

\begin{figure*}[htb!]
	{\centering
		\begin{subfigure}[]
			{\centering{\includegraphics[width = 6.5cm, height = 4.5 cm]{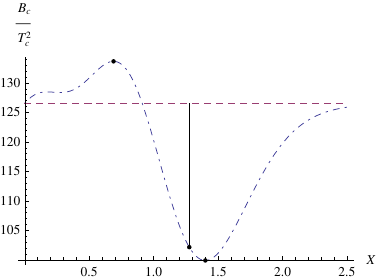}
					\label{pic-2-a}}}
		\end{subfigure}
		~
		\begin{subfigure}[]
			{\centering{\includegraphics[width = 6.5cm, height = 4.5 cm]{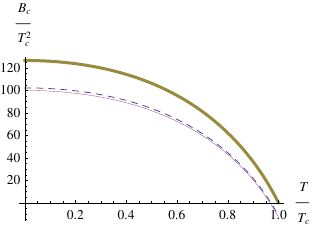}
					\label{pic-2-b}}}
	\end{subfigure}}
	\caption{In Figure \ref{pic-2-a} we  plot scaled $B_c$ against $\Lambda$ for zero value of black hole temperature $T$. Here, the dotdashed curve and the dashed line represent the NC and normal scenarios, respectively. The right side of the vertical line shows the allowed region where $r_+^2 > 4 \theta $. The black points are those for which the values of $\Lambda$ are equal to $0.689$, $1.277$ and $1.401$, respectively. In  Figure \ref{pic-2-b} we  plot scaled $B_c$ against scaled $T$ for  NC and  normal cases. The dashed, continuous and  thick curves  correspond to $\Lambda $  equal to $1.277$, $1.401$ and normal case, respectively.}
	\label{fig-2}
\end{figure*}

Substituting (\ref{neq-46}) into (\ref{B}) and using (\ref{T-hor-rel}, \ref{T-c}) we finally obtain the following expression of the critical value of the magnetic field strength $B_c$, as
\begin{equation}
	B_{c} \simeq \frac{16\pi^{2}T_c^{2}(0)}{9}  \left( 1 - \frac{ 2 \omega^{1/3}}{3 \sqrt{\pi \theta}} e^{-\frac{\omega^{2/3}}{ 4 \theta }} + \frac{\omega}{3 \sqrt{\pi} \ \theta^{3/2}} e^{-\frac{\omega^{2/3}}{ 4 \theta }} \right) \left[ \beta L^2 \sqrt{ 1-\frac{9(3-4z_0) \times g(z_0)}{4(1-z_0)(7-z_0)} \frac{T^4}{T_c^4}} - \frac{(1+2z_0) \times h(z_0)}{(1-z_0)}  \frac{T^2}{T_c^2}  \right] \label{B-c}
\end{equation}
where,
\begin{eqnarray}
	g(z_0) &=& \left[ 1 - \left( \frac{(8 + 25 z_0 - 42 z_0^2)}{3 (3 - 4 z_0) (7 - z_0)} \frac{\omega^{1/3}}{\sqrt{\pi \theta}} - \frac{(1 + 2 z_0)(1 - 3 z_0)}{12 (3 - 4 z_0) (7-z_0)} \frac{\omega}{\sqrt{\pi} \  \theta^{3/2}}  + \frac{5 (1 + 2 z_0)(1 - z_0)}{24 (3 - 4 z_0) (7 - z_0)} \frac{\omega^{5/3}}{\sqrt{\pi} \  \theta^{5/2}} \right) e^{-\frac{\omega^{2/3}}{ 4 \theta }} \right], \nonumber\\
	h(z_0) &=& \left[1 + \left( \frac{(2 - z_0)}{(1 + 2 z_0)} \frac{\omega^{1/3}}{\sqrt{\pi \theta}}  - \frac{(1 + z_0)}{4 (1 + 2 z_0)} \frac{\omega}{\sqrt{\pi} \  \theta^{3/2}} - \frac{(1 - z_0)}{8 (1 + 2 z_0)} \frac{\omega^{5/3}}{\sqrt{\pi} \  \theta^{5/2}} \right) e^{-\frac{\omega^{2/3}}{ 4 \theta }} \right].\nonumber
\end{eqnarray}
From the above expression (\ref{B-c}) it is clear that the strength of the critical magnetic field $B_c$ is affected significantly due to the NC contributions. The Gaussian form of non-commutativity not only affects the coefficient of the dimensionless measure of the temperature $T / T_c$ through $g(z_0)$ and $h(z_0)$, it also forces  the magnitude of $B_c$ to  fluctuate due to the appearance of the term $\left( 1 - \frac{ 2 \omega^{1/3}}{3 \sqrt{\pi \theta}} e^{-\frac{\omega^{2/3}}{ 4 \theta }} + \frac{\omega}{3 \sqrt{\pi} \ \theta^{3/2}} e^{-\frac{\omega^{2/3}}{ 4 \theta }} \right)$ on the right hand side of (\ref{B-c}). It is expected that the former effects will modify the nature of the plot in a qualitative way (in comparison to the normal case), whereas the latter effect will introduce an NC correction to the magnitude of $B_c$ at each point. In order to see how the NC effect affects the magnitude of $B_c$ for the different black hole masses, in Figure \ref{pic-2-a} we have plotted the critical magnetic strength $B_c$ against the variable $\Lambda$ (defined as before by $\Lambda= \frac{\omega^{1/3}}{ 2 \sqrt{\theta}}$) for vanishing temperature $T/ T_c = 0$. In Figure \ref{pic-2-b} we have further plotted the scaled critical magnetic field $B_c/T_C^2$ against the scaled critical temperature $T/T_c$ for those black holes for which the values of $\Lambda$ differ most  from the normal case.

An interesting point to note here is the following: the Black Hole mass $M$ appears explicitly in the NC corrections. However, this is a generic feature of NC corrections and we have already encountered a similar phenomenon in \cite{sg} where the appearance of the probe particle mass term tends to affect the equivalence principle in General Relativity. Essentially the NC parameter $\theta $ introduces a new mass scale and to counter it the existing mass parameters of the system begin to appear in the results. Below we elaborate briefly on the Figures \ref{fig-1} and \ref{fig-2}.

In Figure \ref{fig-1} we have shown the effects of non-commutativity on the value of the condensation operator $O_2$ for different values of black hole temperature $T$. From  Figure \ref{pic-1-a}, it is clear that the NC effect increases the coefficient $\lambda$ of the function $T_c^2 \sqrt{1 - \frac{T}{T_c}}$ near the minimum allowed value of $\Lambda = 1.277$. $\lambda$ is maximum if the ratio between the black hole mass and the NC parameter is equal to $1.366$. However, if the mass of the black holes is large compared to the NC effect, its value stabilizes to the normal case. This result shows that the NC effect increases the value of the condensation operator for a black hole if its mass is roughly of the same order as  the NC parameter. From  Figure \ref{pic-1-b} one can easily see the change of the condensation operator for different black hole mass.

In Figure \ref{fig-2} we have shown the change of the critical magnetic field $B_c$ for different black hole masses as well as for black hole temperature $T$. From Figure \ref{pic-2-a} it is clear that the critical value of the magnetic field is lower than the normal case near the minimum allowed mass of the black hole. Its value first decreases for larger black holes and become minimum if the ratio between the mass of the black hole and NC parameter is equal to $1.401$. However, for more larger black holes $B_c$ increases and stabilizes to the value of the normal case. This result shows that in this NC scenario the critical value of the magnetic field actually reduces from normal case. From Figure \ref{pic-2-b} one can clearly see this result. One can further notice that the nature of the curves does not change very much, whereas the magnitude changes more significantly. For this reason, it is crucial to note that in this NC scenario the critical magnetic field $B_c$ vanishes before the temperature $T$ reaches the critical value $T_c$.

Finally, we are in a position to compare and comment  on our observations as to how the NC effects tend to make the holographic superconductor less stable, with the behavior of a conventional superconductor living in NC space, as studied in \cite{twist}, in a purely condensed matter scenario. It has been observed earlier \cite{xiao}  that  the Berry curvature of the band structure can lead to NC generalized phase space in  quantum mechanics for  electrons in certain materials. In \cite{twist} the authors consider a form of NC space that is induced from  the Berry curvature that appears from  time reversal symmetry  breaking (as in some ferromagnetic materials) or spatial inversion symmetry breaking (as in in GaAs). This form of NC space, known as the Moyal plane, is inspired from the Seiberg-Witten structure of NC \cite{string}.  Quantum field theories on the Moyal plane follow a new form of  quantization, known as  twisted quantization \cite{chai} leading to a modified) commutation relations of the basic oscillator algebra for creation and annihilation operators. (See \cite{bal} for some interesting  consequences.) The authors of  \cite{twist} have explicitly constructed Cooper pairs of electrons using the deformed or twisted oscillator algebra and have shown that {\it{the NC effect reduces the gap energy}} thereby making the superconductor more fragile. Although mathematically the NC structures of \cite{twist} and our model is different still both essentially leads to a minimum area in configuration space either from uncertainty principle \cite{string} or from smearing \cite{nico2}. In this perspective the qualitative agreement between the behaviors of our NC generalized holographic superconductor and the superconductor residing on the Moyal plane is indeed encouraging. It is also apparent the NC in the bulk can generate a NC in the boundary theory via AdS-CFT connection. Further analysis can reveal a mapping between the two forms of NC parameters which can shed new light on the holographic superconductors since note that the (effective) NC parameter used in \cite{twist} is a physical parameter induced from outside, and not imposed in an ad-hoc way from outside. As a more ambitious program, it might be possible to relate the bulk NC to any material property of the high $T_c$-superconductors which makes them more stable, (eg. it is well-known that $Bi_2CaSr_2Cu_2O_9$ is more stable than $YBa_2Cu_3O_7$) or to some external influence (eg. stability of the cuprate structures in high $T_c$-superconductors depends on external pressure and other parameters).

\section{Conclusions and future prospects}

We have studied an extended form of AdS-CFT correspondence by introducing non-commutative geometry effects, which is the generalization of the charged black hole in the AdS background. First, we have studied the stability and laws of the black hole thermodynamics through the Euclidean formalism of black hole thermodynamics. Indeed, we have shown that the entropy area law of the black hole thermodynamics is satisfied in this NC scenario and the system is stable since the heat capacity is positive just outside the event horizon. Secondly, we have calculated the stress-energy tensor of the boundary theory. In fact, we have shown that the trace of the stress-energy tensor vanishes at the asymptotic boundary, though it has trace anomaly at the bulk.  

In a previous paper \cite{sou} we have studied  such NC effects on holographic superconductors   in the presence of electric field only. In the present work exploiting an analytic formulation we have further studied the NC effects on holographic superconductors in the presence of a constant external magnetic field. We have restricted ourselves to the probe limit and to the lowest non-trivial order in NC parameter. Our findings are summarized below.

We find that the basic features of AdS-CFT correspondence and holographic superconductors remain intact under non-commutative geometry, (at least to the order of approximations considered here). But the critical parameters undergo NC corrections in an interesting and non-trivial way. Apart from the modifications in the numerical values of the parameters, more interestingly, the parameters depend on the NC parameter $\theta $ in a non-linear way and reaches a peak value depending on a dimensionless combination of $\theta $ and the black hole mass $M$. The appearance of mass parameters is in itself interesting and nicely corroborated with (some of our) previous observations on NC corrections in a completely different context.

Our main conclusion is that non-commutative effects tend to work against black hole hair formation which in turn shows its negative effect on holographic superconductors in two ways: it reduces the critical magnetic field value and furthermore the  critical magnetic field vanishes slightly before the temperature reaches its critical value. A major point of interest is that we have been able to show that there is qualitative agreement on the stability property of NC generalized holographic superconductors (in the AdS-CFT correspondence scenario) and the conventional superconductors living on NC space or the Moyal plane  (in pure condensed matter scenario).

Let us finally list some of the  open problems along which one can proceed further.\\
It will be worthwhile to consider a top-down approach where noncommutativity is identified with the anti-symmetric tensor $B_{\mu\nu}$ in the low energy limit of string theory  in the   Seiberg-Witten form \cite{string}. The corresponding boundary theory describing   a holographic superconductor can be studied.

We expect  non-commutative effects to significantly affect the properties of  holographic superconductors in a qualitative way once we go beyond the probe limit and include back reaction. The other area we wish to investigate in the near future is the transport properties where non-commutative effects can have non-trivial consequences.\\





{\bf{Acknowledgements:}} We would like to thank Dr. Dibakar Roychowdhury for helpful discussions.\\ 



\end{document}